\documentclass[twocolumn,showpacs,preprintnumbers,amsmath,amssymb]{revtex4}

\begin{document}
\draft

\title{Larmor precession time, Wigner delay time 
and the local density of states in a quantum wire}

\author{P. Singha Deo}
\address{S. N. Bose National Centre for Basic Sciences, JD Block,
Sector III, Salt Lake City, Kolkata 98, India.}
\date{\today}

\begin{abstract}
Buttiker-Thomas-Pretre (BTP) [Z. Phys B
{\bf 94}, 133 (1994)] proposed that
the concepts behind the Larmor precession time tell us that it is 
possible to define exactly the local density of states (LDOS) 
in terms of the
scattering matrix. However,
we take into account evanescent modes and
show that for an impurity in a quantum wire, 
this is in principle not exactly true.
We also prove that the Wigner delay time gives correct
superluminal times at the Fano resonances, in spite of the
fact that the stationary phase approximation is not valid there.
\end{abstract}


\maketitle

\section{Introduction}

Scattering phase shift in a scattering problem carry a lot of physical
informations and are as important as scattering cross section or
scattered intensity. However, phase shifts were always difficult
to measure directly, until very recently \cite{yac,cer}.
Elastic scattering plays
a prominent role in mesoscopic systems, wherein the inelastic processes
are quenched by reducing the temperature and the sample size \cite{dat}.
Transport processes and thermodynamic properties of such samples
can be formulated in terms of elastic scattering \cite{btp, das}. 
Resonances in
these systems, due to elastic processes, will be mainly Fano resonances
and Breit-Wigner resonance is a special case of Fano resonance
\cite{bag,deo98}.
Impurities in such systems act like point scatterers and delta
function like potentials are studied in this regard \cite{bag, boe}.
Such a negative delta function potential can create Fano resonance. 


\section{Scattering Phase Shift at Fano Resonance}

In this section we shall review some early works that lead us to
study the present problem. Readers who are already familiar in this
area can skip this section.

Phase shifts for Fano resonance was first studied with respect to
parity effect.
Electronic states in a one-dimensional (1D) 
ring with an Aharonov-Bohm flux piercing the
ring exhibit the parity effect according to which if the magnetization
of the ring with $N$ electrons is diamagnetic (or paramagnetic) then
the magnetization with $N+1$ electrons is paramagnetic (or diamagnetic).
Leggett conjectured that this is true even in presence of interaction
between electrons as well as defects or
disorder in the ring \cite{leg}. Now suppose
another quantum system $S$ is coupled to the 1D ring $R$ such that the 
states of $S$ can leak into $R$ and become flux dependent. 
Ref. \cite{deo96} 
shows that
the states of the combined system (S+R) does not always have the 
property of reversing
the magnetization with the addition of a single electron.
It was found that the electrons
in the ring undergo usual phase changes associated with their
quantum mechanical motion. These phase changes are i) Aharonov-Bohm
phase, ii) statistical phase, which means electrons being fermion,
acquire a phase change of $\pi$ when they cross each other and 
iii) relative phase change due to the wave like property of the
electrons, that depend on their wave vector or their kinetic energy.
It was also shown \cite{deo96} that apart from these phase
changes, there are discontinuous phase changes by $\pi$, at the
zeroes of the Fano resonances (say at energy $E_0$)
that will be there when the ring (with
the system S attached to it)
is severed at a point and two leads are attached to the two broken
ends \cite{deo96}.
If Fermi energy of $N$ body state is below $E_0$,
and that of the $N+1$ body state is above $E_0$,
then the magnetization of the $N$ body state and the $N+1$ body
state is the same, and otherwise opposite. Thus Leggett's arguments
can be generalized to systems that are of the form $S+R$, with the
conclusion that this new discontinuous phase change is a new phase
different from the ones mentioned in i), ii) or iii) \cite{deo96}.
When S and R are not 1D systems but becomes
quasi-1D (Q1D), even then this provides a general theory
to understand coupled systems \cite{sre}.

In an experiment \cite{yac}, a quantum dot was coupled to a ring
(S+R is now a dot+ring)
and the conductance oscillations of the system with an Aharonov-Bohm
flux was measured. Yeyati and Buttiker \cite{yey} 
try to interpret the conductance oscillations
in terms of the flux dependence (or magnetization) of the electronic
levels of the combined system (dot + ring) using the Friedel
sum rule (FSR). The phase change at
the resonances of the dot could be roughly understood but the
phase change between the resonances of the dot could not be understood.
Other works tried to assign this in between resonance
phase change to processes like charge
decapture (the system throws away unit charge) \cite{sil}.
Ref. \cite{jay96} and \cite{deo98}
predicted that the resonances are actually Fano
resonances and the phase change between the resonances
can be explained by the new phase 
at the transmission zeroes and it will be an abrupt drop. It was
confirmed in a later experiment that this phase
drop occur over an energy scale, that is much smaller
than any energy scale present in the system \cite{hei}. 
Finite width of the leads and evanescent modes
has to be considered to explain \cite{deo98} the
abrupt phase changes occurring between each consecutive resonances.
Ref. \cite{deo98}
considers 
there is a pole between the zeroes and that
is where a charge is captured and the resonance phase changes smoothly
by $\pi$ in agreement with FSR. But the discontinuous phase drops 
being a new kind of phase,
does not have anything to do with charge capture
or decapture.
Later on this was 
proved to be true whenever we have time reversal
symmetry \cite{lee,tan}. 
Also the fact that simple 1D calculations will not explain
the phase drop between each two consecutive
resonances was shown \cite{yey00}.
Rigorous experimental verification has also emerged in favor of
Fano resonances in quantum dots \cite{kob}.

In fact unitarity is also required to produce such
discontinuous phase changes \cite{deoc}. It was subsequently seen that
when an unitary channel continuously evolves into a non-unitarity 
channel, then along with it, the transmission zeroes evolve
into minima (difference with Breit-Wigner resonance is that
the minima has a complex zero, that is for a complex value
of incident energy, transmission amplitude is zero), and
discontinuous phase drops evolve into continuous phase drops
\cite{swa1}. As an
example, one can consider a two channel quantum wire with a delta
potential (see Fig. 1) at the middle of the quantum wire. 
Since the two channels
of the quantum wire are opposite parity states like the ground state
and the first excited state of a quantum well, the two channels are
decoupled. If an electron is incident from the left
in the first channel then
amplitude of
transition to the second channel i.e., $t_{12}$ (state 1 to state 4)
or $r_{12}$ (state 1 to state 2)
is zero, making the first channel preserve unitarity. 
Whenever $|t_{11}|^2$ has a zero,
the phase of $t_{11}$ drops (or decreases) discontinuously by $\pi$. Now 
if the delta potential is shifted slightly from the center
of the quantum wire, then parity is no longer a good quantum state
and $t_{12}$ etc are no longer zero. So by continuously displacing
the delta potential from the center, one can destroy
the individual unitarity of each channel and thus
make the discontinuous phase
drops continuously evolve into fast continuous phase drops.
These phase drops can be seen in Fig. 2 at an energy
$Ea \approx 85$.
Even dephasing can result in the loss of unitarity \cite{deoc} and can
explain the small width of the phase drops
observed in the experiment.
For the two propagating channel case,
\begin{eqnarray}
 {\partial \theta_f \over \partial E} &=&
 {1 \over 2 \pi}\left[|r_{11}|^2 {\partial arg(r_{11})
 \over \partial E}
 + |r_{22}|^2 {\partial arg(r_{22}) \over \partial E}\right.\nonumber\\
 & & \left. + 4 |r_{12}|^2 {\partial arg(r_{12}) \over \partial E}
 + |t_{11}|^2 {\partial arg(t_{11}) \over \partial E}\right.\nonumber \\
 & & \left. + |t_{22}|^2 {\partial arg(t_{22}) \over \partial E}\right]
 \end{eqnarray}
 Here $\theta_f$ stands for Friedel phase.
For the delta potential at the center of the wire, the 
discontinuous drops in
$arg(t_{11})$ or $arg(t_{22})$ occur when $|t_{11}|^2$=0 and 
$|t_{22}|^2$=0 and so it is obvious from Eq. 1 that there are 
no drops in $\theta_f$. In that case,
obviously the drops in $arg(t_{11})$ or  $arg(t_{22})$
do not have anything to do with $\theta_f$ and hence
charge decapture. But for the $\delta$ potential off center,
since $|t_{11}|^2$ or $|t_{22}|^2$ are not zero when
there are drops in their arguments,
it is not obvious that
these drops do not have anything to do with charge decapture.
However only after all the simplification,
one finds that \cite{swa1}
 \begin{equation}
 {\partial \theta_f \over \partial E}= {\partial 
 arg(r_{\alpha \beta}) \over \partial E}
 \end{equation}
 That means $\theta_f$ only depends on the phase of reflection
amplitudes and hence is completely independent of the phase
drops in the transmission amplitudes. 
So once again the phase drops do not imply a drop in $\theta_f$ or
charge decapture \cite{swa1}.

One also knows from earlier known results that FSR in mesoscopic
systems is valid only
in semi-classical regimes. And so Fano resonance being a purely
quantum interference phenomenon, one would expect larger violations
of FSR at the Fano resonances. Instead what Refs. \cite{swa2} shows is that
FSR is exact at the energy corresponding to the Fano resonance,
but there are large violations away from this energy. Refs. \cite{swa2}
also explained that the correctness of FSR at the Fano resonance
is due to the fact that there is a quasi bound state here,
and hence the self energy due to the leads become minimum and hence
its energy derivative becomes 0. This is obviously true for
any potential that supports Fano resonances.
The exactness of the FSR at the Fano resonance ($Ea \approx 85$) 
and the deviation away from it is
also shown in Fig. 2 (compare the dotted and dashed curves) 
for the two propagating channel case. The results are the
same for any number of propagating channels. The regime around $Ea \approx 
85$ is a purely quantum regime like the regime $Ea<40$. However, in the regime
around $Ea \approx 85$, which is the Fano resonance regime, FSR holds good
unlike the regime $Ea<40$, which is unexpected. $40< Ea < 80$ is the 
semiclassical regime and violation of FSR here is also unexpected.


\section{The Problem}

Apart from the DOS we know that time scales associated with
a particle crossing a quantum mechanical potential can also be determined
from the scattering phase shifts.
For example, in the the stationary phase approximation (i.e.,
phase shifts do not strongly depend on energy) ${d \over dE}
arg(t_{\alpha \beta})$ gives the Wigner delay time (WDT)
for the particle to
be transmitted from state $\alpha$ to state $\beta$.
A negative slope in $arg(t_{\alpha \alpha})$, like that in the solid
and dash-dotted curve at $Ea \approx 85$ in Fig. 2, 
means super-luminality,
i.e., the particle can travel faster than light across the potential,
according to the WDT.
However, once again, since Fano resonance is a quantum
interference effect, dispersion will be very strong and stationary
phase approximation will not hold good at the Fano resonances, and
one cannot be sure if these negative slopes actually correspond to
super-luminality.
To be sure of superluminality, established theories say
that one should see if
we get negative delay times from the Buttiker-Thomas-Pretre (BTP) 
formalism.
They proposed that the Larmor precession
time can be determined exactly from the scattering matrix
and give the correct local delay times in all regimes \cite{btp}
and delay time can be determined by integrating the local delay times.
It gives
 \begin{equation}
 \nu(\alpha,E,r,\beta)=-{1 \over 4 \pi i}
 Tr[S^\dagger_{\alpha \beta} {\delta S_{\alpha \beta} \over
 \delta V(r)} - HC]
 \end{equation}
 where $\nu(\alpha,E,r,\beta)$ is proportional
to the time spent by the particle (i.e., delay time) going from state
$\alpha$ to state $\beta$, while encountering the potential at $r$.
If we sum
it up for all $\alpha$ and $\beta$ then it should
give the LDOS at $r$, exactly \cite{btp, tes}.
Originally, it was derived by considering the effect of a small
magnetic field on the outgoing spin wave function. But more generally,
for any potential, to obtain the LDOS at $r$, we have to create
a $\delta$ function potential like local perturbation at $r$ and
see the change in the scattering matrix $S$ of the entire system.
Thus the delta function potential, apart from representing
point defects,  is also very ideal to study BTP
formula and delay time. 

Given the fact that the scattering phase shift at the Fano resonance
often violate established theories, 
is the BTP formula correct in Q1D? And secondly, is there superluminality
at Fano resonance? WDT suggests that there could be limitless
superluminality but can we trust WDT at the Fano resonance?


\section{The Scattering Solution}

The scattering problem is defined in Fig. 1.
For this system in Fig. 1 \cite{bag},
 \begin{equation}
 t_{\alpha \beta}=r_{\alpha \beta}=
 -{i \Gamma_{\alpha \beta} \over 2d \sqrt{k_\alpha k_\beta}}
 \end{equation}
 where for the transmissions, $\alpha \ne \beta$ and
 \begin{equation}
 t_{\alpha \alpha}= r_{\alpha \alpha}+1
 \end{equation}
 \begin{equation}
 d=1+\sum_e {\Gamma_{ee} \over 2 \kappa_e} + i 
 \sum_\alpha{\Gamma_{\alpha \alpha}\over 2k_ \alpha}
 \end{equation}
 For a quantum wire with hard wall confinement,
$\Gamma_{mn}=(2m_e\gamma/\hbar^2)sin[{m\pi \over w} (y_i + w/2)]
sin[{n\pi \over w} (y_i + w/2)]$,
$m_e$ is electron mass, $k_n=\sqrt{(2m_e/\hbar^2)(E-E_n)}$,
$\kappa_n=\sqrt{(2m_e/\hbar^2)(E_n-E)}$, $E_n=(\hbar^2/2m_e)
(n^2 \pi^2/w^2)$,
$E$ is incident energy, $\sum_e$ is sum over
all the evanescent modes.


{\it Explicit calculations of LDOS:} We derive
LDOS explicitly from the internal wavefunctions for unit incident
flux by using \cite{tan}
 \begin{equation}
 \rho(E,x=0,y=y_i)=\sum_\alpha {2 \over hv_\alpha }|\psi_{\alpha}(x=0,y=y_i)|^2
 \end{equation}
Where $\psi_{\alpha}(x,y)$ is the wavefunction when incident electron
is in the $\alpha th$ channel. It can be taken from Ref. \cite{bag}
and can be further simplified to give
 \begin{equation}
 \rho(E,0,y_i)=
 \sum_\alpha
 {2\over hv_\alpha}
 |\sum_{m} t_{\alpha m}
 sin({m\pi \over w}(y_i+w/2))|^2
 \end{equation}
 where $v_\alpha = \hbar k_\alpha /m_e$ and \cite{swa2},
 \begin{equation}
 t_{\alpha e}=
 \frac{-\frac{\Gamma_{\alpha e}}{2\kappa_e}}{1+\sum_e\frac{\Gamma_{ee}}
 {2\kappa_e}+i \sum_\alpha  \frac{\Gamma_{\alpha \alpha}}{2k_\alpha}}
 \end{equation}
 In Eq. 8, sum over $\alpha$ runs over the propagating modes 
only while that for $m$ is for all modes.

{\it LDOS from BTP formula:}
We consider two propagating channels to present our results
but we have checked that the results are the same for
any number of channels including just one channel.
For this case, Eq. 3 after summing over $\alpha$ and $\beta$ gives
 \begin{equation}
 \nu(E,0,y_i)=-{1 \over \pi i}
[r_{11}^* r_{11}' + r_{22}^* r_{22}' + 4 r_{12}^* r_{12}'
   + t_{11}^* t_{11}' + t_{22}^* t_{22}' - HC]
 \end{equation}
 where primes mean derivatives wrt $V(0,y_i)$.

\section{Verification of BTP Formula}

The derivation  of Eq. 3
assumes that a small perturbation to the actual system, allows us to
expand a scattering matrix element as
\begin{eqnarray}
S_{\alpha \beta}^{\pm}(E,V(x_i,y_i) \mp \delta V(x_i,y_i))  =
 S_{\alpha \beta}(E,V(x_i,y_i))\nonumber\\
 \quad\quad \mp \int dx_i' dy_i' [\delta S_{\alpha \beta}(E,V(x_i', y_i'))/
 \delta V(x_i',y_i')] \delta V(x_i',y_i') + \ldots
\end{eqnarray}
If we make a substitution of the type
 \begin{equation}
 -\int dx_i' dy_i' {\delta \over \delta V(x_i',y_i')} \rightarrow
  {d \over dE}
 \end{equation}
 then we derive FSR.
But this substitution being
an approximation, the FSR is also approximate.

First of all, note that the role of evanescent modes
on the derivative of the
scattering matrix elements is that it renormalizes
$\delta \gamma$
according to the following relation.
 \begin{equation}
 \delta \gamma 
 \sum_e^N {g_{ee} \over 2 \kappa_e}= \delta \gamma^/ \quad (say)
 \end{equation}
 where $g_{ee}={2 m_e \over \hbar^2} sin^2[{e \pi \over w}(y_i + w/2)]$.
Note that the sum $\sum_e^N {g_{ee} \over 2 \kappa_e}$ 
is not a converging series.
It diverges as log[N] as $N \rightarrow \infty$ \cite{boe}. 
Here $N$ is the total number of evanescent modes. And so derivative
w.r.t $\gamma$ will not exist for any arbitrary number of modes.
Thus
the expansion required to derive BTP formula
is not defined at all energies
and thus the concept of Larmor precession time fails. That means
LDOS cannot be defined in terms of S matrix.
We shall also show that the
expansion in Eq. 11 is valid at the Fano resonance
and also explicit calculations of LDOS prove the correctness
of BTP formula only at the Fano resonance.
For this note that at the bound state
$\sum_e {g_{ee} \over 2 \kappa_e}=-{1 \over \gamma}$,
irrespective of the number of evanescent modes.
Thus at the Fano resonance, convergence exists and
the expansion in 11 holds good.
Which means that at the Fano resonance, the BTP formula will
give the correct delay time and hence also the LDOS.
We verify this explicitly by numerically calculating the RHS of 8 and 10. 
For example, let us truncate the series at the 3rd term, that is we
consider 2 evanescent modes only. In that case we show in Fig. 3
that the BTP formula is accurate. To calculate the derivative we
have taken $\delta \gamma$=0.001. However this kind of agreement
does not occur for example when we consider 5000 evanescent modes
with $\delta \gamma$=0.001 (see Fig. 4). 
This can be cross checked by using Eq. 13.
Note that now although the two curves
do not coincide with each other exactly, they do coincide exactly
at the energy corresponding to the bound state (i.e.,
where both curves peak). 
Of course one can take
a smaller value of $\delta
\gamma$ and get a better agreement between the two
curves, but then again the same disagreement will be there if more
evanescent modes are considered. Since the sum is not a converging sum,
this will be a never ending story. 
One can see that ${\delta S_{\alpha \beta} \over \delta \gamma}
= {\delta S_{\alpha \beta} \over \delta \gamma '}{\delta \gamma '
\over \delta \gamma} $ do not exist for an infinite number
of evanescent modes as ${\delta \gamma '
\over \delta \gamma} $ diverge as log[$N$].

It is possible to provide further analytical arguments in support
of our result. The RHS of Eq. 10 can be simplified to give \cite{btp}
\begin{equation}
\nu(E,0,y_i)=-{1\over 2 \pi i} \sum_{\alpha \beta} {Tr [S^\dagger_{
\alpha \beta} {\delta S_{\alpha \beta} \over \delta V(0,y_i)}]} 
\end{equation}
or
\begin{equation}
\nu(E,0,y_i)={\delta \over \delta V(0,y_i)}[-{1 \over 2 \pi i}
log Det[S]]
\end{equation}
It was also shown in ref. \cite{swa2}
\begin{equation}
-{1 \over 2 i} log Det[S] = -Arctan{Im[r_{\alpha \beta}]
\over Re[r_{\alpha \beta}]} + constant
\end{equation}
This means that the functional derivative in 15 can be calculated
by expanding the RHS of 16 and considering only the linear term
in $\delta V(0,y_i)$, provided all higher order terms are finite. 
In the regime ($1+ \sum_e {{\Gamma_{ee} \over 
2 \kappa_e}}) > \sum_\alpha {{\Gamma_{\alpha\alpha} \over 2 \kappa_\alpha}}$, this
expansion can be done using (4) and (6), to give
$$Arctan{Im[r_{\alpha \beta}] \over Re[r_{\alpha \beta}]}
={\pi \over 2} - \sum_\alpha{{\Gamma_{\alpha\alpha} \over 2 k_\alpha}}(1+\sum_e
{{\Gamma_{ee} \over 2 \kappa_e}})^{-1} +$$ 
\begin{equation}
{1 \over 3}
(\sum_\alpha{{\Gamma_{\alpha\alpha} \over 2 k_\alpha}})^3
(1+\sum_e
{{\Gamma_{ee} \over 2 \kappa_e}})^{-3} ....
\end{equation}
It can be seen that the coefficients of higher order terms as well
as the coefficients of linear terms in $\delta \gamma$, diverge
as log[N] as well as higher powers of log[N], 
which implies that the functional derivative does not
exist. 

Wang et al \cite{wan} has pointed out some regimes where the BTP
formula is violated due to the lack of gauge invariance. They
also calculated the correction terms in those regimes. However,
they overlooked the violations that can arise due to non-existence
of functional derivatives, and we point it out in this paper.
Only way out is to truncate the number of evanescent modes to keep
the error within acceptable limits.

\section{Generalization}

The scattering matrix $S$ of an extended system
can be written as \cite{dat} $S=S_1 \bigotimes S_2 \bigotimes S_3$,
where $S_2$ is the scattering matrix of an infinitesimal region
at $r$, $S_1$ and $S_3$ are the scattering matrices of the regions to the
left and right of $r$, respectively. Hence $S_2$ is the scattering matrix
of a $\delta$ function potential that we have to further perturb
infinitesimally to see the changes in $S_2$. All the change in $S$
will be due to this change in $S_2$. Thus complications of the BTP
formula will depend on its complications for a $\delta$ function
potential.
If we want to integrate LDOS over an extended region then one has to
take $\delta$ function potential like perturbations at many many places
of the region and sum the changes they produce on the S matrix. The error
will be added.

\section{WDT at Fano Resonance gives correct Superluminality}

We have shown in section 5 that the BTP formula is exact at the Fano
resonance. 
So the LHS in
12, operating on $S_{\alpha \beta}$ will give the correct
delay time as will be observed in an experiment.
The substitution
in 12, is exact at the Fano resonance follows from earlier results
\cite{swa2} that the FSR is exact at Fano resonance. 
Hence the RHS of 12, operating on $S_{\alpha \beta}$ will also give
the correct delay time as can be observed in an experiment.
Hence, in spite of all non-stationary phase behavior, the WDT
will exactly correspond to physical delay times at the Fano resonance
and the negative slopes in Fig. 2 do mean that there is strong
superluminality that can be observed.

\section{Large violations of FSR due to evanescent modes}

We see in Fig. 2 that away from the Fano resonance, there are large
violations in FSR. Once we have explored the BTP formula we can
now analyze the cause of it. First of all we would like to state
that we are dealing with a system that is coupled to reservoirs
(a grand canonical system) where the reservoirs can inject charge
or absorb charge. So charge is not conserved,
$\rho - \rho_0$ can be arbitrary, and it may not be possible to 
relate it to ${d \theta_f \over dE}$ very strictly.
But of course, as in all grand canonical systems, at equilibrium
charge and energy are as good as conserved and we can talk of
Friedel sum rule. 
Eqn 12 suggests that if there is a sample connected to
semi-infite quasi 1D leads then
$${d \theta_f \over dE} = \pi [\rho(E)-\rho_0(E)]_{global}=$$
\begin{equation}
\pi 
\sum_\alpha \frac{2}{hv_\alpha}
 \int_{-\infty}^{\infty} [|\psi(r)|^2 - |\psi_0(r)|^2]dr
\end{equation}
where $r$ represents coordinate. Now the sample that extends
from $-r_s$ to +$r_s$ is a grand canonical system
and one can show that when the leads are sigle channel then
(in the absence of evanescent modes)
$${d \theta_f \over dE} = \pi[\rho(E)- \rho_0(E)] =$$ 
\begin{equation}
\pi \sum_\alpha \frac{2}{hv_\alpha}
 \int_{-r_s}^{r_s} [|\psi(r)|^2 - |\psi_0(r)|^2] dr - {S-S^\dagger \over 4
(E-E_1)}
\end{equation}
${S-S^\dagger \over 4(E-E_1)}$ is the error due to the substitution
in Eq. 12. It depends on parameters like $E_1$ that depends on the
internal details of the potential. For multichannel leads,
all the $E_n$s appear in the correction term. 
This amount is a negligible amount for practical
purposes and the substitution in 12 can still be used without
involving large errors. We show below that in the presence of evanescent
modes, there will be larger errors when we make the substittuion in 12.

Ref. \cite{swa2} did some explicit calculations to show
$$[\rho(E) - \rho_0(E)]_{global} = \sum_\alpha \frac{2}{hv_\alpha}
 \int_{-\infty}^{\infty} propagating \quad  modes $$
\begin{equation}
 + \sum_\alpha \frac{2}{hv_\alpha}
 \int_{-\infty}^{\infty} evanescent \quad modes 
\end{equation}
that gives
 \begin{eqnarray}
 [\rho(E) - \rho_0(E)]_{global}& = & \sum_\alpha \frac{2|{r_{\alpha\alpha}}|}{hv_\alpha}
 \int_{-\infty}^{\infty} dx\hspace{.2cm} Cos(2k_\alpha x+\eta_\alpha)\nonumber\\
 &+&\sum_\alpha \frac{2}{hv_\alpha}\hspace{.2cm}\sum_e\frac{\mid t_{\alpha e}\mid^2}{\kappa_e}
 \hspace{.2cm}.
 \label{eq:sysDOS}
 \end{eqnarray}
 Here $v_\alpha=\hbar k_\alpha/m_e$.
$\sum_\alpha$ denotes sum over all propagating modes and $\sum_e$ denotes sum
over all evanescent modes. $r_{\alpha\alpha}=
|r_{\alpha\alpha}|e^{-\eta_\alpha}$ is the $\alpha$th
diagonal element in the $S$-matrix. $t_{\alpha e}$ is the transition
amplitude from the $\alpha$th propagating mode to the $e$th evanescent
mode.

So now if we follow the scheme of 19 then $\rho(E) - \rho_0(E)$=0
as $r_s \rightarrow 0$.
But if we consider 
 \begin{eqnarray}
 [\rho(E) - \rho_0(E)]& = & 
 \sum_\alpha \frac{2}{hv_\alpha}\hspace{.2cm}\sum_e\frac{\mid t_{\alpha e}\mid^2}{\kappa_e}
 \hspace{.2cm}.
 \end{eqnarray}
then we get the dotted curve in Fig 2. The first term
on the RHS of 21, gives the charge lost to the reservoir.
One can do the integration to find 
$\int_{-\infty}^{\infty} dx Cos(2k_\alpha x+\eta_\alpha)
= \pi Cos(\eta_\alpha) \delta(k_\alpha)$.  So for $k_\alpha >0$,
charge is fully conserved.
The difference in the dotted and
dashed curves in Fig.2,
is the error due to the substitution 12. The error is still $S-S^\dagger
\over 4(E-E_1)$ in case of single channel leads
and this error is due to the substitution in 12. 

Substituion 12 means that change in scattering matrix element
due to infinitesimal increase (decrease) in incident energy, is the
same as change in scattering matrix element due to constant
infinitesimal decrease (increase) in potential over entire space.
And also the constant decrease in potential over entire space can be
integrated as small changes locally and adding up the effect
due to all such local changes.

If we could have chosen a $\delta V(r)$ at $r<-r_s$ that disturbs the
propagating modes at $r$ and does not disturb the evanescent modes
at $r$ then the error in substitution 12 would have been as negligible
as it is in 1D.
This is because although the evanescent modes exist in the leads
they are to be considered as not existing in the leads but existing
in the sample (this is clear from 19, 20, 21, 22). 
Indeed we see that in Eq 6, if we ignore any change in the
second term on the RHS (which is due to evanescent modes)
then the scattering matrix elements look exactly like that
of a delta function potential in 1D. But once we include the
evanescent modes on the RHS of 6,
we get diverging terms
in the scattering matrix if we change $E$ or $V$, and the 
substitution described in the above para involves
much greater error as compared to 1D.

\section{Summary and conclusions}

The BTP formalism is very crucial to understand mesoscopic
transport beyond the Landauer conductance formula, that is
beyond the linear and DC response.
So far it has been verified for simple 1D systems \cite{tes}
or systems where there are localized states completely
decoupled from the leads \cite{tex}.
Violations of BTP formula can also occur in situations
where there is absence
of gauge invariance \cite{wan}.
We have shown that BTP formalism
is not exact in Quasi-one-dimension (Q1D) 
due to the presence of evanescent modes.
These evanescent modes make the scattering matrix
singular in nature and the 
series expansions in Eq. 11.
required to show the equality between $\rho$ and $\nu$ breaks down.
If there are only a few evanescent modes in the system, then the formula
may be acceptable for practical purposes. For extended potentials
and large number of evanescent modes, it may not be practical.
It definitely cannot be used as a definition for DOS. DOS has to be
defined in terms of the internal wave function and Hamiltonian.
Scattering matrix will not contain all the information.
We have also proved that 
in spite of all non-stationary phase effects,
WDT (${d \over dE}
arg(t_{\alpha \alpha})$) 
correctly give the superluminal times and there is superluminality
at the energy where the phase drops of transmission
amplitudes occur (for example at $Ea$=85 in Fig. 2).
Experimentalists have always tried to find situations
wherein they can create larger superluminality. A delta
function potential in a Q1D wire can create limitless
superluminality.
So far, experiments to observe superluminality, only considered systems
with Breit-Wigner type resonances \cite{ste}.

Finally in this paragraph, we also make some comments on possible
future research, that are not related to the theme of this work.
Features of superluminality in case of Fano type resonances
should be explored experimentally. 
Also as it seems that semi-classical formulas like WDT and FSR
are valid at Fano resonances, in spite of strongly energy dependent
scattering phase shifts, it is possible that an incident wave
packet is not dispersed by scattering at Fano resonance. Parts of
it will be transmitted to different channels without shape distortions.
Rather, if wavepackets are simultaneously incident on the scatterer
from different channels, then after scattering the outgoing wavepackets 
in different channels will be similar to the incoming wavepackets.
Their centroids may remain unchanged before and after scattering.
This normally happens for solitons and very rare in quantum mechanics.

\centerline{Figure captions}

Fig. 1. Here we show a quantum wire of width $w$ with a delta
function potential $V(x,y)=\gamma \delta(x)  \delta(y-y_{i})$ situated
at $\times$. We consider scattering
effects when the incident electron is from the left. The sub-bands on the
left of the impurity is denoted as 1 for the first mode and 2 for the second mode.
Similarly the sub-bands on the right of the impurity is denoted as 3 for
the first mode
and 4 for the second mode.

Fig. 2. The dotted curve gives G=$\pi (\rho(E) - \rho_0(E))$,
where $\rho(E)$ is DOS at energy E in presence of the scatterer
and $\rho_0(E)$  is DOS at energy E in absence of the sctterer.
The system considered is in Fig. 1, and G is
in units of $g=(m_e w^2/\hbar^2)$.
The dashed curve gives $Q=d \theta_f/dE$ in units of $q=g$.
The solid curve gives $arg(t_{11})$ in radians, displaced on the
y-axis by 0.1.
The dash-dotted curve gives $arg(t_{22})$ in radians, displaced on the
y-axis by 0.8.
We have taken $\gamma = -15$, $y_i$=0.45
and 2 propagating modes along with 2 evanescent modes.
The x-axis is energy in units of
$a=w^2$.

Fig. 3. The system under consideration is
shown in Fig. 1. with two propagating modes.
Solid line gives $H=\rho(E,0,y_i)$, that LDOS calculated
from internal wavefunction, in units of $h=(2mw/\hbar^2)$.
Dashed line gives $J=\nu(E,0,y_i)$, i.e., LDOS calculated
using BTP formula, in units of $j=h$.
All parameters are the same as that considered in Fig. 2.
The x-axis is energy in units of $a=w^2$.

Fig. 4. Here we consider 5000 evanescent modes and $\gamma =-1.5$.
Everything else is the same as in Fig. 3.

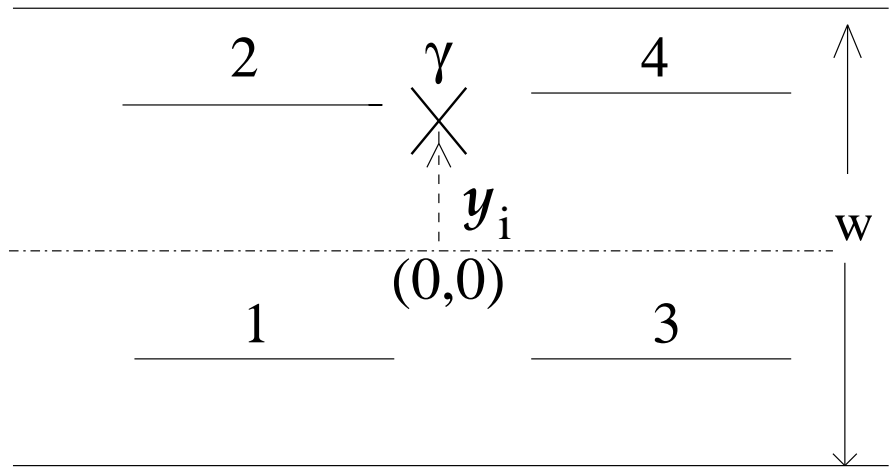
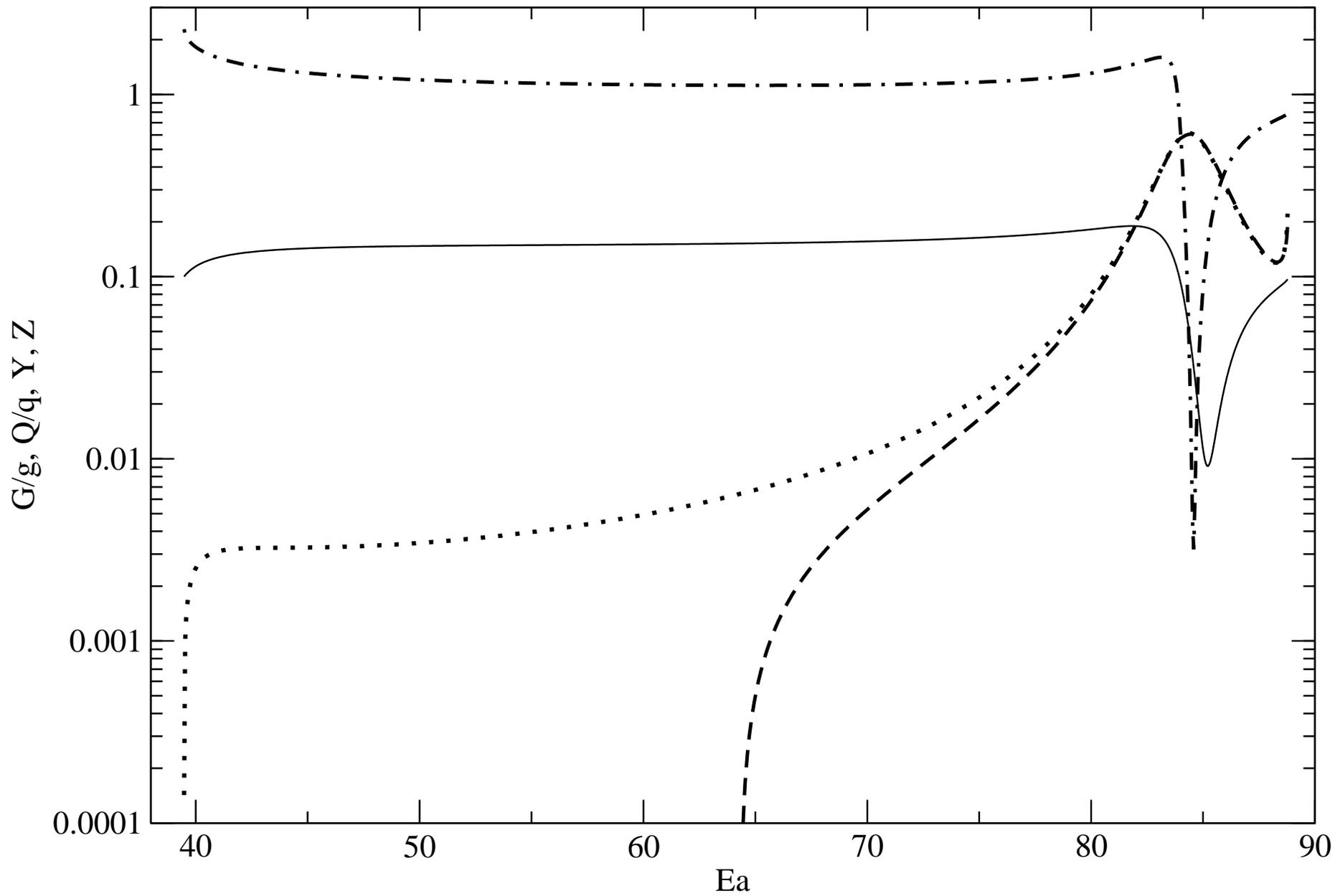
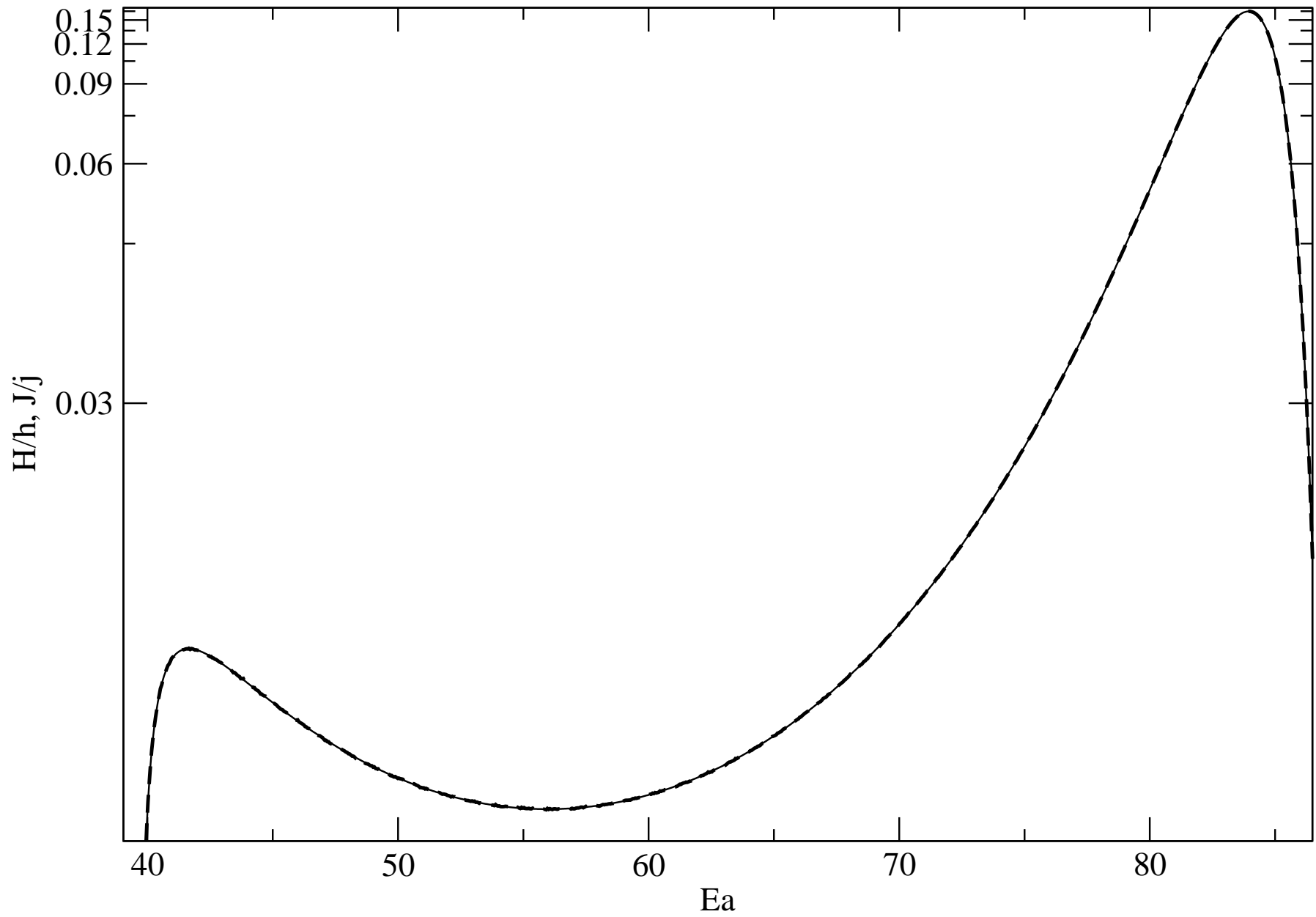
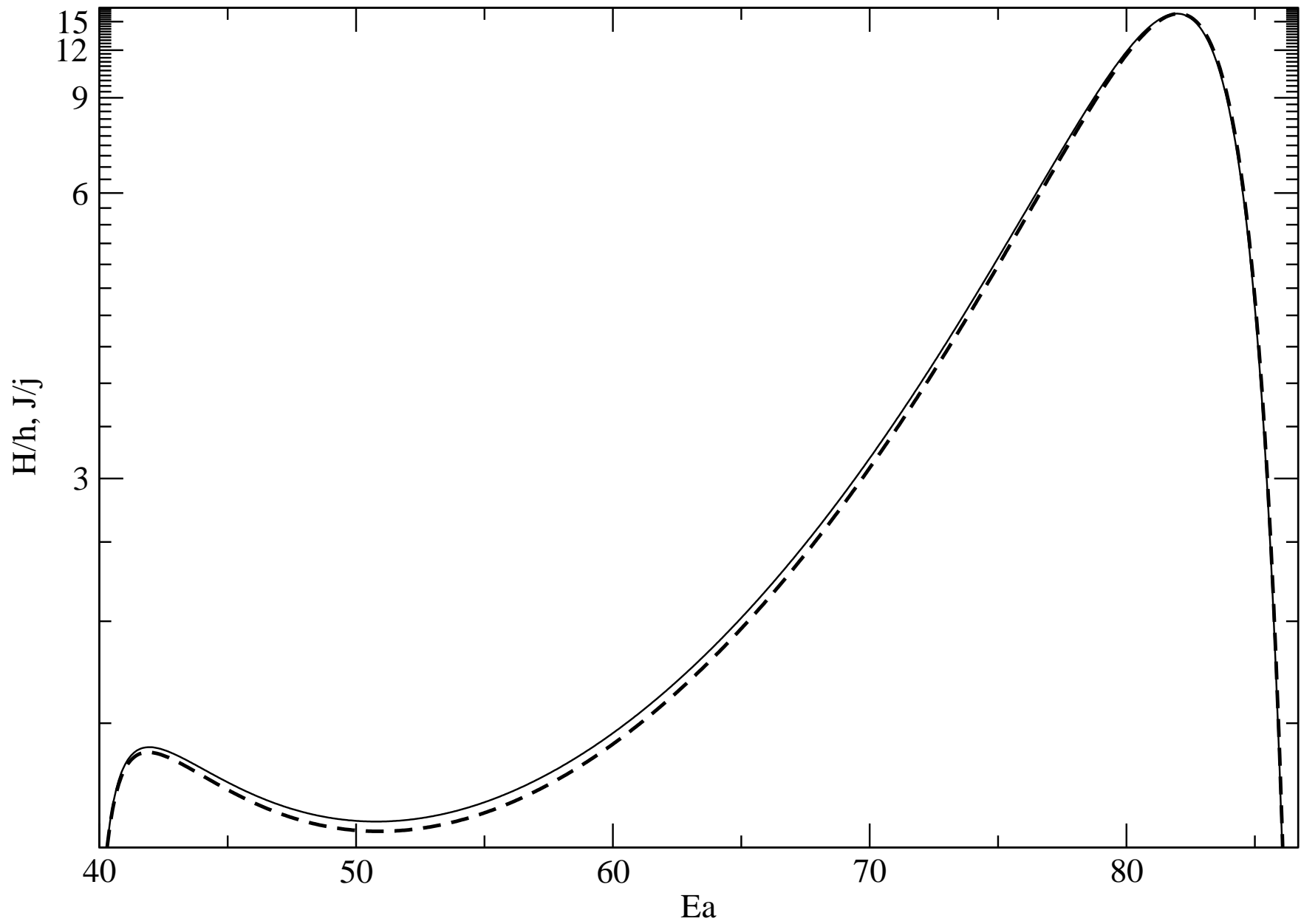
\end{document}